\documentclass[structabstract]{aa}

\usepackage{epsfig}
\usepackage{graphicx}
\usepackage[latin1]{inputenc}
\usepackage{txfonts}
\usepackage{rotating}
\usepackage{natbib}

\begin{document}

   \title{Galactic neutron stars}

   \subtitle{I. Space and velocity distributions in the disk and in the halo}

   \author{N. Sartore
          \inst{1},
          E. Ripamonti
          \inst{1,2},
          A. Treves
          \inst{1}
          \and
          R. Turolla
          \inst{3,4}
          }

   \institute{Dipartimento di Fisica e Matematica, Università dell'Insubria,
              via Valleggio 11, 22100, Como, Italy
              \email{nicola.sartore@gmail.com}
         \and
             Dipartimento di Fisica, Università di Milano Bicocca,
             Piazza delle Scienza 3, 20126, Milano, Italy
         \and
         		 Dipartimento di Fisica, Università di Padova,
         		 via Marzolo 8, 35131, Padova, Italy
         \and
         	   Mullard Space Science Laboratory, University College London, Holmbury St. Mary, Dorking, Surrey, RH5 6NT, UK
             }

   \date{Received ...; accepted ...}

 \abstract
	{}
	{Neutron stars (NSs) produced in the Milky Way are supposedly ten to the eighth - ten to the ninth, of which only $\sim\,2\,\times\,10^{3}$ are observed. Constraining the phase space distribution of NSs may help to characterize the yet undetected population of stellar remnants.}
	{We perform Monte Carlo simulations of NS orbits, under different assumptions concerning the Galactic potential and the distribution of progenitors and birth velocities.
   We study the resulting phase space distributions, focusing on the statistical properties of the NS populations in the disk and in the solar neighbourhood.}
	{It is shown that $\sim\,80$ percent of NSs are in bound orbits. The fraction of NSs located in a disk of radius 20 kpc and width 0.4 kpc is $\lesssim\,20$ percent. Therefore the majority of NSs populate the halo. Fits for the surface density of the disk, the distribution of heights on the Galactic plane and the velocity distribution of the disk, are given. We also provide sky maps of the projected number density in heliocentric Galactic coordinates (\textit{l}, \textit{b}). Our results are compared with previous ones reported in the literature.}
	{Obvious applications of our modelling are in the revisiting of accretion luminosities of old isolated NSs, the issue of the observability of the nearest NS and the NS optical depth for microlensing events. These will be the scope of further studies.}


\maketitle

\section{Introduction}
Neutron stars are born during core-collapse of massive ($M \geq 8 M_{\odot}$) stars (type Ib, Ic and II supernovae, herafter SNe) or, less frequently, by accretion-induced collapse of white dwarves that reached the Chandrasekhar's limit (type Ia SNe).
A rough estimate of the total number of NSs generated in the Milky Way (MW) can be obtained from the present-day core-collapse SN rate, $\beta_{NS}$, which is of the order of a few per century \citep{D06} and assuming $\beta_{NS}$ constant during the lifetime of the Galaxy ($\sim\,10$ Gyr). Hence the total number of NSs created in the MW lies between $10^{8}$ and $10^{9}$. NSs may then represent a non negligible fraction of the Galactic stellar content.

Up to now only $\sim\,2\,\times\,10^{3}$ NSs have been observed, the majority of which as isolated radio pulsars (PSRs) with ages far shorter ($\lesssim\,100$ Myr) than the MW lifetime.
Older NSs have been detected only when recycled in binary systems by mass and angular momentum transfer from a companion star, thus becoming millisecond pulsars (see \citealt{L08} and references therein).
Isolated old NSs (ONSs) have not been identified so far because, once
their energy reservoir, both thermal and rotational is exhausted, they
are pretty close to being invisible.
As a consequence little is known about their physical and statistical properties.

On the other hand the expected phase-space distribution of ONSs can be constrained by means of population synthesis models once a realistic set of initial conditions is given.
Population synthesis studies of Galactic NS have been performed by many authors in the past. \cite{H90} and \citet[hereafter H90 and P90 respectively]{Pa90} studied the orbits of Galactic NSs, looking for a possible link with gamma-ray bursts.
These studies differed in their assumptions. In particular H90 assumed a Gaussian distribution centered at 200 $\rm{km\,s}^{-1}$ for the distribution of NS birth velocities, while P90 adopted the distribution of \cite{L82} which gives a different (higher) weight to the low velocity tail of the distribution.

NS orbits in the Galactic gravitational potential were also investigated by \cite{BR}, \cite{BM}, \cite{Z95}, \cite{PP98} and \citet[hereafter BR, BM, Z95, PP98 and P00 respectively]{P00}  in order to constrain the number of nearby NSs accreting from the interstellar medium (ISM, see \citealt{T00} and references therein).
BR, BM and Z95 adopted initial conditions similar to those of P90 except the distribution of birth velocities which, following \cite{NO}, was assumed to be Maxwellian with a dispersion of 60 $\rm{km\,s}^{-1}$.

\cite{P00} explored the observability of accreting ONSs for a wide range of initial mean velocities, between 0 and 550 $\,\rm{km\,s}^{-1}$, assuming a Maxwellian distribution.
The paucity of observed accretors\footnote{No sources have been positively identified so far.} in the ROSAT catalogue (\citealt{NT99}) led to the conclusion that NSs are born with average velocities of at least $200\,\rm{km\,s}^{-1}$.

This is confirmed by observations of known young NSs.
PSRs show in fact spatial velocities of several hundreds $\rm{km\,s}^{-1}$, i.e. of the same order of the escape velocity from the MW (see e.g. \citealt{L97}, \citealt{A02}, \citealt{B03}, \citealt{H05} and \citealt{F06}; hereafter L97, A02, B03, H05 and F06 respectively).
Some PSRs exhibit velocities in excess of 1000 $\rm{km\,s}^{-1}$.
A striking example is PSR B1508+55: the proper motion and parallax measurements obtained from radio observations points to a transverse velocity of $\sim\,1083\,\rm{km\,s}^{-1}$ (\citealt{C05}).

Similar high values of the velocity have been inferred also for objects belonging to other classes of isolated NSs.
Thanks to Chandra observations, \cite{HB06} estimated a velocity of $\sim$ 1100 $\rm{km\,s}^{-1}$ for the central compact object RX J0822-4300.
Recently \cite{M09} measured the proper motion of one of the ROSAT radio-quiet, thermally emitting NSs (the Magnificent Seven) and found a value of the 3D velocity of $600\,-\,1000\,\rm{km\,s}^{-1}$.
This is not uncommon in PSRs and hence they concluded that the velocity distribution of the Magnificent Seven is not statistically different from that of normal radio pulsars.

The origin of such high velocities is not at all clear.
An asymmetric SN explosion is considered one possible explanation (e.g. \citealt{S70}, \citealt{DC87}).
Also the effects of binary disruption (e.g. \citealt{B61}, \citealt{IT96}) may contribute to the observed velocities.
Recently it has been proposed that the fastest NSs are the remnants of runaway progenitors expelled via N-body interactions from the dense core of young star clusters \citep{GGP08}.

If all classes of isolated NSs share the same typical birth velocities, no matter how these are achieved, a large fraction of these objects can escape the potential well of the MW in a relatively short time.
This fact has consequeces of all observable NS populations.
In this paper we focus on the effect of high birth velocity and likely evaporation on the still elusive population of ONS.
Constraining the expected phase space distribution is in fact crucial to define suitable strategies for thier detection.

Based on recent estimates of the birth velocity distribution, in this paper we reconsider the dynamics of isolated NSs.
We perform integration of stellar orbits using our new code PSYCO (Population SYnthesis of Compact Objects), developed for this purpose.
In Section \ref{method} we describe the ingredients of the simulation, i.e. the gravitational potential of the Milky Way and the distributions of progenitors and birth velocities.
We present the results of the simulation in Section \ref{results}. In particular we investigate the statistical properties of the NS population in the Galactic disk and at the solar circle. We fit the surface density of the disk and the average height distribution and compute the surface and volume densities in the solar vicinity.
We fit also the velocity distribution in the disk, both with respect to the Galactic center and the local rest frame of the ISM.
We discuss our results and their possible implications in Section \ref{discussion}.
The results of this work will constitute the base for further studies on the observability of ONSs.

\section{METHOD}\label{method}

We follow an approach similar to P90.
Initial conditions (position, velocity) are taken randomly from the selected distributions and assigned to each synthetic NS by means of a Monte Carlo procedure.

\subsection{Distribution of progenitors}

The initial positions of NSs in the Galaxy are defined in a galactocentric cylindrical coordinates system $(R,\phi,z)$, where the \textit{z} axis corresponds to the axis of rotation of the MW.
These initial positions reflect the distribution of NS progenitors: according to \citet[hereafter B00]{B00}, formation of massive stars is currently concentrated in a annular region which follows the distribution of molecular hydrogen.
However, to explore the effects of different initial conditions on the current phase-space configuration of NSs, we choose four possible radial distributions of progenitors from the literature.

P90 adopted an exponential probability distribution, based of the observed surface brightness\footnote{In th J band.} of face-on Sc galaxies \citep{vdK}

\begin{equation}
	p(R)\,dR = a_{R}\frac{R}{R_{exp}^{2}} \exp\Big(-\frac{R}{R_{exp}}\Big)\,dR\,,
 \end{equation}

\noindent where $p(R)dR$ is the probability that a NS is born between $R$ and $R+dR$, $R_{exp}\,=\,4.5$ kpc and $a_{R}\,=\,1.0683$.

B00 obtained the already mentioned radial distribution of starforming regions in the MW from the combined far infrared (FIR) and millimetric emission produced by clusters of massive stars embedded in ultra-compact HII regions.
The FIR (surface) luminosity, $\rho(R)$, has a Gaussian shaped rise until it reaches a maximum at $\sim 4.7$ kpc (FWHM of 2.38 kpc), then it decays exponentially, with a scale-length of 1.78 kpc, towards the outer part of the MW, approximately following the distribution of neutral hydrogen.
The radial birth probability is then obtained from the FIR luminosity from the equation

\begin{equation}\label{rho-p}
		p(R)\,dR =  \frac{R\,\rho(R)\,dR}{\int_{0}^{\infty}\,R\,\rho(R)\,dR}\,.
	\end{equation}

Another possible distribution of NS progenitors can be obtained from the surface density of Galactic SN remnants \cite[hereafter CB98]{CB98}

\begin{equation}
	\centering
		\rho(R) = \Big(\frac{R}{R_{0}}\Big)^{\alpha}\exp\left[-\frac{\beta(R-R_{0})}{R_{0}}\right]\,,
	\end{equation}

\noindent where $\alpha = 2$ and $\beta = 3.53$ and $R_{0}=8.5$ kpc.
The corresponding radial probability density is again obtained from equation (\ref{rho-p}).

The fourth radial distribution adopted has been proposed by F06

\begin{equation}
	\centering
		p(R) = \frac{1}{\sqrt{2\pi}\sigma}\exp\left[-\frac{(R-R_{peak})^{2}}{2\sigma^{2}}\right]\,,
	\end{equation}

\noindent where $R_{peak} = 7.04$ kpc and $\sigma = 1.83$ kpc.
This distribution has been extrapolated from the observed PSR distribution found by \cite{YK}.
Finally for all models we assume that NSs can be born from 0 to 20 kpc.

It is our opinion that the distribution proposed by P90, in spite of being obtained from observations of external galaxies, may better represent the long term star formation history of the MW.
The other models are based on the present-day distribution of population I objects, which could have been rather different in past epochs (see for example Chiappini et at. 2001).
The models of B00, CB90 and F06 are probably better suited for population studies of young/middle-aged NSs (PSRs, magnetars etc.).

\begin{figure}
		\includegraphics[width=0.47\textwidth]{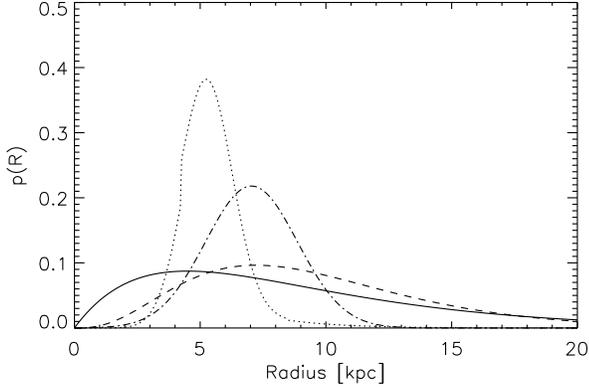}
	\label{fig:Radial_distributions}
	\caption{Normalized radial probability distribution of NSs progenitors.
	P90 (solid line), B00 (dotted), CB98 (dashed) and F06 (dot-dashed).}
\end{figure}

\subsubsection{Spiral arms and initial distribution of heights}
Massive stars are located in the spiral arms of the MW (they are indeed the ideal tracers of the spiral structure), thus we model spiral arms in the distribution of NS progenitors adopting the same prescription of F06, i.e., NS progenitors are distributed along four logarithmic spirals, each spiral described by the equation

\begin{equation}\label{arms}
	\phi(R) = k\,\ln(R/R_{*}) + \phi_{0}\,.
\end{equation}

\begin{figure}
	\centering
		\includegraphics[width=0.45\textwidth]{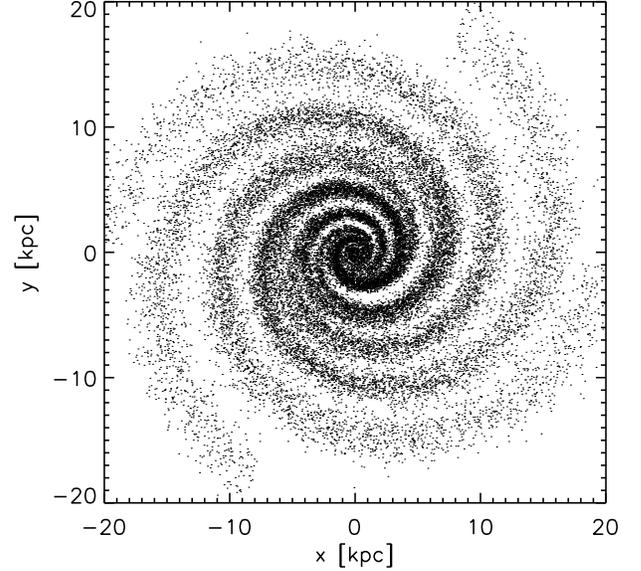}
	\caption{Initial positions of NSs - Radial distribution from P90. The position of the Sun is (8.5, 0.0).}
	\label{fig:init_pos_RA}
\end{figure}

The values of the parameters \textit{k}, $R_{*}$ and $\phi_{0}$ for each spiral are given in Table \ref{tab:spirals}.
Actually, equation \ref{arms} describes the position of arm centroids.
A more realistic distribution can be obtained if the positions of progenitors are scattered, both in the radial and azimuthal directions, around these centroids (Fig. \ref{fig:init_pos_RA}).
Details on how the scatter is added to the initial positions of NSs can be found in F06.

\begin{table}
\centering
\caption{Parameters of the spiral arms.}
	\begin{tabular}{l c c c c c}
\hline
	Arm & k & $R_{*}$ & $\phi_{0}$ \\
	& & [kpc] & [radians] \\
\hline
\\
	Norma & 4.25 & 3.48 & 1.57 \\
	Carina-Sagittarius & 4.25 & 3.48 & 4.71 \\
	Perseus & 4.89 & 4.90 & 4.09 \\
	Crux-Scutum & 4.89 & 4.90 & 0.95 \\
\hline
	\end{tabular}
	\label{tab:spirals}
\end{table}

The thickness of the starforming region is few tens of parsecs (B00, \citealt{M01}).
However, as P90 and \cite{SH04} have pointed out, the long term dynamical behavior of a NS population is insensitive to the scale height of its progenitors (see also \citealt{KH09}).
Following their results we assume that all NSs are born on the Galactic plane ($z=0$).

\subsection{Distribution of birth velocities}\label{kicks}

The true form of the distribution of birth velocities is still a hotly debated issue and few constrains exist (see F06 for an exhaustive discussion).
HP97, L97 and H05 proposed a Maxwellian distribution

\begin{equation}
		p(v) = \sqrt{\frac{2}{\pi}} \frac{v^{2}}{\sigma^{3}}\exp\Big( -\frac{v^{2}}{2\,\sigma^{2}} \Big)\,.
	\end{equation}

\noindent Alternatively, \cite{F98}, \cite{CC}, A02 and B03 proposed a bimodal distribution

\begin{equation}
	p(v) = \sqrt{\frac{2}{\pi}}\,v^{2}\,\Big[ \frac{w}{\sigma_{1}^{3}}\exp\Big( -\frac{v^{2}}{2\,\sigma_{1}^{2}} \Big) +
   \frac{1-w}{\sigma_{2}^{3}}\exp\Big(-\frac{v^{2}}{2\,\sigma_{2}^{2}} \Big) \Big]\,.
	\end{equation}

\noindent where \textit{w} is the relative weight of the two components of the distribution.

Using the same PSR sample of B03, F06 explored, together with Maxwellian and bimodal models, other possible distribution functions like the double-sided exponential

\begin{equation}
		p(v_{i}) = \frac{1}{2\,v_{exp}} \exp\Big(-\frac{\left|v_{i}\right|}{v_{exp}}\Big)\,,
	\end{equation}

\noindent where $v_{i}$ represents a single component of the velocity and $v_{exp}$ is a characteristic velocity; the Lorentzian

\begin{equation}
		p(v_{i}) = \frac{1}{\pi\,\gamma\,\Big(1\,+\,\big(v_{i}^2/\gamma^{2}\big)\Big)}\,,
	\end{equation}

\noindent where $\gamma$ is a scale parameter defining the half-width at half maximum, and the distribution proposed by \cite{L82} and adopted by P90

\begin{equation}
		p(v) = \frac{4}{\pi\,v_{*}\,\Big(1 + \big(v/v_{*}\big)^{2}\Big)^{2}}\,,
	\end{equation}

\noindent where again \textit{v} represents the tridimensional velocity.
F06 concluded that the Maxwellian model is less favored.
On the other hand they disfavour also the bimodal distribution and prefere instead single parameter models, pointing out that the bimodality found by other authors arises if thse alternative single parameter models are not investigated.

To explore the effects of the birth velocities on the final phase-space distribution of NSs, we adopt the Maxwellian model of H05 as well as four of the models proposed by F06, i.e. the bimodal, the double-sided exponential, the lorentzian and that of P90.
From here on we refer to these models as H05, F06B, F06E, F06L and F06P respectively.

The value of the mean tridimensional velocity for each distribution is calculated numerically from simulated velocity vectors (Table \ref{tab:kdist}).
All the velocity distributions refer to the Local Standard of Rest (LSR) of NS progenitors.
Thus, the true 3D velocity of a neutron star with respect to the Galactic Reference Frame (GRF) is the vector sum of the birth velocity and the circular velocity at the birthplace, $\textbf{v} = \textbf{v}_{birth} + \textbf{v}_{circ}$.

\begin{figure}
		\includegraphics[width=0.47\textwidth]{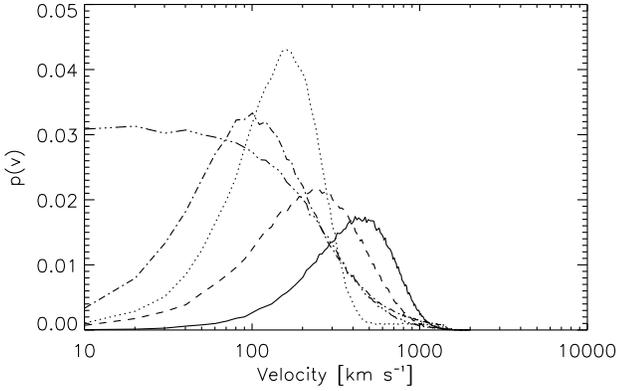}
	\caption{Differential velocity distributions obtained from simulated velocity vectors. H05 (solid), F06DG (dotted), F06E (dashed), F06L (dot-dashed) and F06P (triple dot-dashed).}
	\label{fig:kdist}
\end{figure}

\begin{table}
\centering
	\caption{Velocity distribution models.}
				\begin{tabular}{c c c}
\hline
Model & Parameters & $<\!v\!>$ \\
 			& 					 & [$\rm{km\,s}^{-1}$] \\
\hline
H05 & $\sigma = 265\,\rm{km\,s}^{-1}$ & 420 \\
F06B	& $\sigma_{1}=160\,\rm{km\,s}^{-1}$ & 335 \\
					& $\sigma_{2}=780\,\rm{km\,s}^{-1}$ & \\
					& $w=0.9$ & \\
F06E & $v_{exp} = 180\,\rm{km\,s}^{-1}$ & 400 \\
F06L & $\gamma=100\,\rm{km\,s}^{-1}$& 447 \\
F06P & $v_{*}=560\,\rm{km\,s}^{-1}$& 331 \\
\hline
		\end{tabular}
		\label{tab:kdist}
\end{table}

\subsection{Gravitational potential}\label{gpot}
Once the initial conditions have been assigned, the motion of NSs is described by the equation

\begin{equation}\label{motion}
		\ddot{\textbf{r}} = -\nabla\Phi\,,
	\end{equation}
 	
\noindent where $\textbf{r}=\textbf{r}(R,\phi,z)$ is the position of the NS and $\Phi$ is the gravitational potential of the MW. We adopt the same 3-component model of \citet[hereafter S07]{S07}

\begin{equation}
	\Phi = \Phi_{B} + \Phi_{D} + \Phi_{H}\,,
 \end{equation}

\noindent where $\Phi_{B}$, $\Phi_{D}$ and $\Phi_{H}$ represent the bulge, disk and halo contributions respectively.

The gravitational potential of the bulge is \citep{He90}

\begin{equation}
	\Phi_{B} = -\frac{GM_{B}}{r + r_{B}}\,,
 \end{equation}

\noindent where $M_{B}\,=1.6\times10^{10}\,M_{\odot}$ and $r_{B}\,=0.6$ kpc are respectively the mass and the scale radius of the bulge and $r\,=\!\sqrt{R^{2} + z^{2}}$ is the distance from the GC.

The disk potential has instead the following form \citep{MN75}

\begin{equation}
	\Phi_{D} = -\frac{GM_{D}}{\sqrt{\left\{R^{2} + \left[R_{D} + \sqrt{z_{D}^2 + z^{2}}\right]^{2}\right\}}}\,,
 \end{equation}

\noindent where $M_{D}\,=5\times10^{10}\,M_{\odot}$ is the mass of the disk and the $R_{D}\,=\,4$ kpc and $z_{D}\,=\,0.3$ kpc are respectively the scale length and scale height of the disk.

Finally, the potential of the halo is \citep{NFW}

\begin{equation}
	\Phi_{H} = -\frac{4\pi\,G\rho_{s}r_{vir}^{3}}{c^{3}r}\log\Big(1 + \frac{c\,r}{r_{vir}}\Big)\,,
 \end{equation}

\noindent where

\begin{equation}
\centering
\rho_{s} = \frac{\rho_{cr} \Omega_{0} \delta_{th}}{3} \frac{c^{3}}{\ln(1+c) - c/(1+c)}
\end{equation}

\noindent is the characteristic density, \textit{c} is the concentration parameter, $r_{vir}$ is the virial radius and $\rho_{cr}$ is the critical density of the Universe.

The parameters of potential are the same of S07, except for the concentration parameter \textit{c} and the virial radius $r_{vir}$ (19.2 and 274 kpc respectively), which were adjusted to match the IAU standard values for the distance of the Sun from the Galactic center, $R_{0}\,=\,8.5$ kpc, and the circular velocity at the solar circle, $v_{circ}(R_{0})\,=\,220\,\rm{km\,s}^{-1}$, together with the escape velocity from the MW at the same distance, $v_{esc}(R_{0}) \simeq 544\,\rm{km\,s}^{-1}$ (S07).
The corresponding value of the virial mass, $M_{vir}$, is $\sim\,10^{12}\,M_{\odot}$.

Very recently \cite{Re09} gave a new estimate of the circular velocity, $v_{circ}(R_{0})\simeq254\,\rm{km\,s}^{-1}$ with $R_{0}=8.4\,\rm{kpc}$, meaning that the MW may be more massive that previously thought.
To asses the effect of the enhanced mass of the Galaxy on NS orbits, we choose a further set of parameters for the potential: the masses of the bulge and disk are increased by a factor $(254/220)^{2}$, i.e. the ratio of the squared circular velocities in the two cases.
For the halo, the concentration parameter \textit{c} remains the same while the virial radius $r_{vir}$ is in this case 332 kpc, which yields an 80 percent increase of the virial mass, $M_{vir}\,\sim\,1.8\times10^{12}\,M_{\odot}$.

We note that in the model where $v_{circ}(R_{0})\,=\,254\,\rm{km\,s}^{-1}$ we get $v_{esc}\,=\,664\,\rm{km\,s}^{-1}$.
This is higher than the central value (544 km/s) estimated by S07; however, it is not far from their 90\% upper limit (608 $\rm{km\,s}^{-1}$), especially when we consider that $v_{esc}$ was obtained by assuming $v_{circ}\,=\,220\,\rm{km\,s}^{-1}$, and that modifying such assumption introduces further uncertainty in its determination (Smith, private communication).

\begin{figure}[htbp]
	\includegraphics[width=0.47\textwidth]{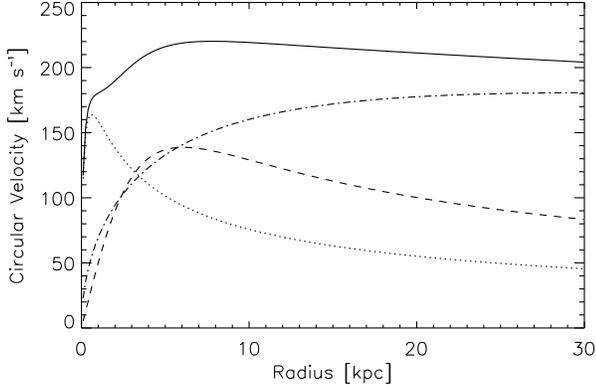}
	\label{fig:vesc}
	\caption{Rotation curve for our Milky Way model (solid). Dotted, dashed and dot-dashed represent the bulge, disk and halo contributions respectively.}
\end{figure}

\begin{figure}[htbp]
  \includegraphics[width=0.47\textwidth]{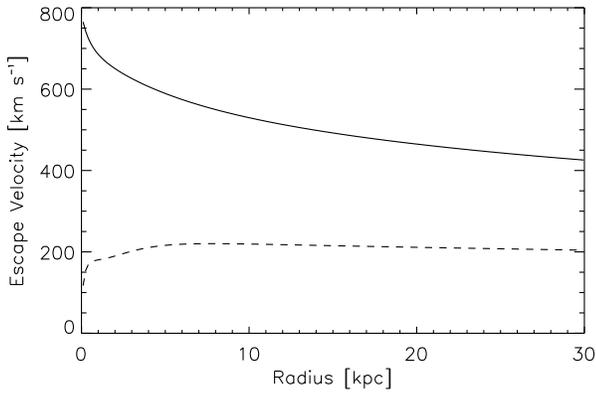}
	\label{fig:vesc}
	\caption{Escape velocity on the Galactic plane. The circular velocity (dashed) is plotted for comparison}
\end{figure}

\subsection{Orbit Integration}
We calculate the orbits of $150 000$ NSs for each model (Table \ref{tab:models}), assuming that NSs are born at constant rate during the whole MW lifetime (10 Gyr).
Hence the age of each NS is selected randomly from an uniform distribution.
The orbit of each NS is then calculated via numerical integration of the system of equations (\ref{motion}), for a time corresponding to its assigned age.
The axial symmetry of the potential implies conservation of angular momentum with respect to the axis of rotation of the MW.
This allows to reduce the number of equations in (\ref{motion}) to four
\\
\begin{eqnarray}\label{motion2}
	\frac{dR}{dt} & = & v_{R}\,, \nonumber \\
	\frac{dz}{dt} & = & v_{z}\,, \nonumber \\
	\frac{dv_{R}}{dt} & = & \frac{\partial\Phi}{\partial\!R}\,+\,\frac{j_{z}^{2}}{R^{3}}\,, \\
	\frac{dv_{z}}{dt} & = & \frac{\partial\Phi}{\partial\!z}\,, \nonumber
\end{eqnarray}

\noindent where $j_{z}$ is the angular momentum with respect to the \textit{z} axis.
Integration of equations (\ref{motion2}) is performed with a 4th order Runge-Kutta algorithm (e.g. \citealt{P92}) with customized adaptive stepsize.
The relative accuracy of integrations is kept below $10^{-6}$ using the energy integral E as reference, i.e. $(\delta\!E/E)\,\leq\,10^{-6}$, where

\begin{equation}\label{eint}
		E = \frac{v^{2}}{2}\,+\,\Phi(\textbf{r})\,.
	\end{equation}

To limit the computation time\footnote{The CPU time for a typical run is about 1 day.} and avoid lockups of the code, all NSs reaching $0.1$ kpc from the Galactic center are discarded.
The fraction of NSs traveling to within 0.1 kpc from the Galactic center is less than 1 percent in each run.

\begin{table}
\centering
	\caption{Models for initial conditions. (*) denotes models with updated potential.}
		\begin{tabular}{c c c c c c}
\hline
 & \multicolumn{5}{c}{Birth velocity distr.} \\
Spatial & H05 & F06B & F06E & F06L & F06P \\
distr. & & & & & \\
\hline
P90 & 1A & 1B & 1C & 1D & 1E \\
B00 & 2A & 2B & 2C & 2D & 2E \\
CB98 & 3A & 3B & 3C & 3D & 3E \\
F06 & 4A & 4B & 4C & 4D & 4E \\
P90 & 1A* & 1B* & 1C* & 1D* & 1E* \\
\hline
		\end{tabular}
	\label{tab:models}
\end{table}

\section{RESULTS}\label{results}

\begin{figure}
	\centering
		\includegraphics[width=0.47\textwidth]{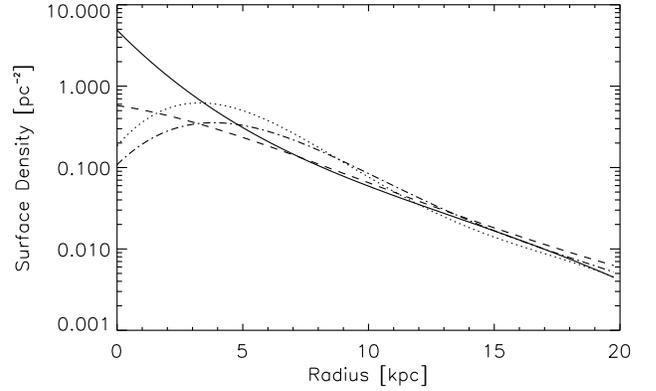}
	\caption{Surface density of NSs in the disk, obtained from best fit parameters ($N_{star}=10^{9}$). Models 1B (solid line), 2B (dotted), 3B (dashed) and 4B (dot-dashed).}
	\label{fig:surfs}
\end{figure}

Our calculations show that the statistical properties of NSs are affected mostly by the distribution of birth velocities while the effects of different distributions of progenitors are less prominent.
For this reason we focus on results of models 1A to 1E, i.e. with the distribution of progenitors of P90.
Results of models differing only for the distribution of progenitors are quite similar, the only substantial difference is the shape of the surface density (Fig. \ref{fig:surfs}): in fact, in models based on the P90 progenitor distribution, the density peaks at the center ($R_{peak}=0$) whereas for other models the density peaks off center ($2 \leq R_{peak} \leq 6$ kpc).

\subsection{Fraction of bound neutron stars}

We first compute the fraction of NSs in bound orbits, $f_{bound}$.
Neglecting all those processes that could alter its energetic state (e.g. two body interactions), the final fate of a NS is known once its initial position and velocity are fixed.
A NS star is bound when its initial velocity is lower than the escape velocity at the birthplace, $v_{i} < v_{esc}(\textbf{r})$, with

\begin{equation}
	\centering
		v_{esc}(\textbf{r}_{i}) = \sqrt{-2\Phi(\textbf{r}_{i})}\,,
	\end{equation}

\noindent where $\textbf{r}_{i}$ is the position of the newborn NS. Thus

\begin{equation}
	\centering
		f_{bound} = \frac{N(v<v_{esc})}{N_{star}}\,.
	\end{equation}

The retention fraction is $\sim$ 0.7 for models 1A, 1C and 1D, while for models 1B and 1E, $f_{bound}\,\sim$ 0.9 and 0.8 respectively (Table \ref{tab:disk}).

\subsection{Distribution of heights}

From here on our results are obtained rescaling $N_{star}$ from $150 000$ to $10^{9}$, which is our reference value for the total number of NSs produced in the MW.

\begin{figure}
	\centering
		\includegraphics[width=0.47\textwidth]{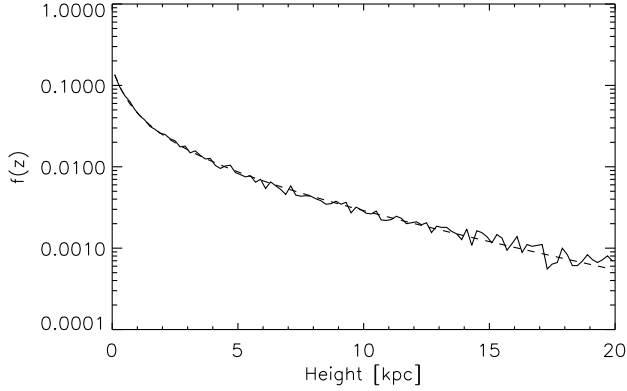}
	\caption{Distribution of heights \textit{f(z)} - Model 1A. Dashed line represents the fitting function.}
	\label{fig:zdist}
\end{figure}

We study the distribution of NSs, $f(z)$, as a function of the height on the Galactic plane.
We adopt a logistic function

\begin{equation}
	\centering
		f(z) = \frac{1}{(b_{0}b_{1}^{z} + b_{2})}\,,
	\end{equation}
	
\noindent as fitting function (see e.g. Fig. \ref{fig:zdist}).
From these fits we estimate the average half density half thickness $z_{1/2}$ of the disk (Table \ref{tab:disk}).
The values of the coefficients of the fit for each model, together with the corresponding maximum error, are listed in the Appendix (Table \ref{tab:fit_z}).
The half density half thickness shows substantial variations from model to model, going from 100 to $\sim\,400$ pc for models 1A to 1D: for model 1E in particular, $z_{1/2}$ is $\sim$ 30 pc, i.e. roughly an order of magnitude smaller than those obtained from other models.
We will return on this fact later.

\subsection{Neutron stars in the disk}\label{res_disk}

Here we define the Galactic disk as the cylindrical volume with radius 20 kpc and height 0.4 kpc (i.e $R\leq 20$ kpc and $|z| \leq 0.2$ kpc).
The fraction of NSs that reside in the disk, $f_{disk}$, goes from $\sim\,0.05$ to $\sim\,0.20$.
Hence the majority of NSs born in the MW populate the halo (Table \ref{tab:disk}).

We fit the logarithmic surface density of the disk adopting a fourth order polynomial as fitting function

\begin{equation}
	\centering
		\rm{log}\,\Sigma(\it{R}) = a_{0} + a_{1}R + a_{2}R^{2} + a_{3}R^{3} + a_{4}R^{4}\,.
	\end{equation}
	
\noindent The accuracy of these fits is always better than 5 percent. The values of the coefficients $a_{j}$ are listed in the Appendix (Table \ref{tab:fit_surf}).

\begin{figure}
	\centering
		\includegraphics[width=0.45\textwidth]{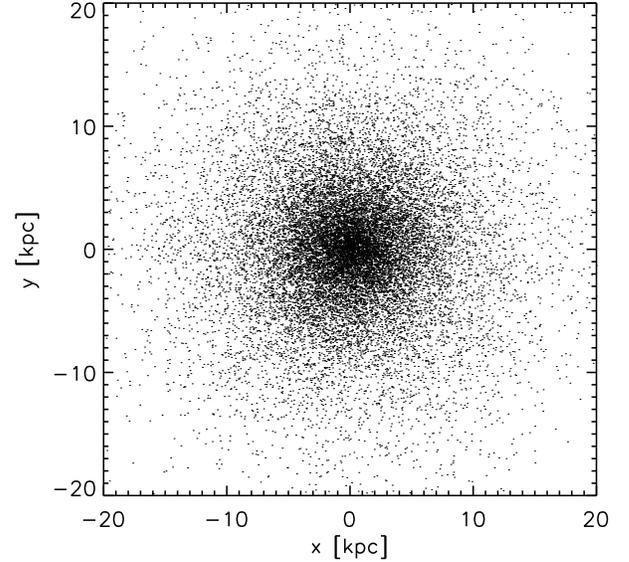}
	\caption{Final distribution of NSs in the disk - Model 1E*.}
	\label{fig:pos_xy}
\end{figure}

We made a visual check of the final distribution of NSs in the disk, looking for traces of the spiral arms.
We found no evidence of spiral structure in the evolved distribution (compare Figs. \ref{fig:pos_xy} and \ref{fig:init_pos_RA}).

\begin{figure}
	\centering
		\includegraphics[width=0.47\textwidth]{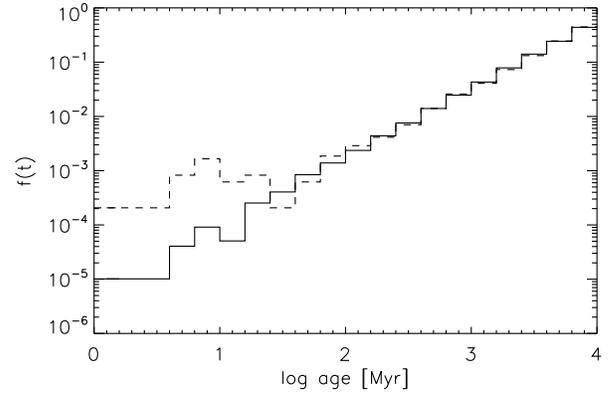}
	\caption{Distribution of ages - Model 1A. Solid an dashed lines represent global and disk populations respectively.}
	\label{fig:age}
\end{figure}

The hypothesis of constant formation rate yields an average age of $\sim\,5$ Gyr, with young/middle-aged ($< 10$ Myr) NSs representing only $\sim\,0.1$ percent of the whole population.
As expected this value is higher inside the disk, by a factor $\sim$ 10, since NSs are born there, and did not have enough time to run away.
However the excess of young NSs (Fig. \ref{fig:age}) does not alter the mean age in the disk, which is also $\sim 5$ Gyr.

\begin{table*}
	\centering
	\caption{Statistical properties of NSs in the disk.}		
		\begin{tabular}{c c c c c c c c c c c}
\hline
Model & 1A & 1B & 1C & 1D & 1E & 1A* & 1B* & 1C* & 1D* & 1E* \\
\hline
\\
$f_{bound}$
& 0.70 & 0.88 & 0.72 & 0.71 & 0.79
& 0.84 & 0.91 & 0.82 & 0.77 & 0.85 \\
\\
$f_{disk}$
& 0.05 & 0.11 & 0.10 & 0.12 & 0.19
& 0.06 & 0.13 & 0.12 & 0.16 & 0.21 \\
\\
$z_{1/2}$
& 367 & 225 & 164 & 100 & 33
& 345 & 192 & 149 & 80 & 28 \\
$[\rm{pc}]$ \\
\\
$<\!v\!>$
& 230 & 220 & 215 & 213 & 213
& 262 & 250 & 249 & 245 & 245 \\
$[\rm{km\,s}^{-1}]$ \\
\\
$<\!v^{LSR}\!>$
& 180 & 146 & 199 & 164 & 82
& 199 & 156 & 216 & 176 & 89 \\
$[\rm{km\,s}^{-1}]$ \\
\\
$f_{50}$
& 0.003 & 0.001 & 0.003 & 0.002 & $\ll\,0.001$
& 0.002 & $<\,0.001$ & 0.002 & $<\,0.001$ & $\ll\,0.001$ \\
\\
$f_{50}^{LSR}$
& 0.019 & 0.060 & 0.033 & 0.074 & 0.437
& 0.014 & 0.039 & 0.029 & 0.068 & 0.379 \\
\\
\hline
	\end{tabular}
		\label{tab:disk}
\end{table*}

\subsubsection{Mean velocities}\label{res_vel}

The mean velocity of NSs in the disk is roughly the same for all models, $<\!v\!>\,\sim\,210\,-\,230\,\rm{km\,s}^{-1}$ in the GRF while in the LSR the mean velocity is lower, $<\!v^{LSR}\!>$ $\sim\,150\,-\,190\,\rm{km\,s}^{-1}$.
An exception are models based on the distribution F06P, which show mean velocities in the LSR of $\sim 80\,\rm{km\,s}^{-1}$.
This fact can be easily explained: in the F06P model low birth velocities have higher probability (see Fig. \ref{fig:kdist}) and thus the main contributor to the velocity of the star is the circular velocity, $\textbf{v}_{birth} + \textbf{v}_{circ} \simeq \textbf{v}_{circ}$.

The low velocity in the LSR implies also that NSs can not move too far away from the disk and that is the reason why, for models F06P, the scale height is considerably lower than in other models.

Following Z95, we fit the cumulative velocity distribution of NSs, both with respect to the GRF and the LSR, with the following function

\begin{equation}
	 G(v) = \frac{\big(v/v_{0}\big)^{m}}{1+\big({v/v_{0}}\big)^{n}}\,,
\end{equation}

\noindent where a $v_{0}$ is a characteristic velocity.
Fit values for $v_{0}$, \textit{m} and \textit{n} are listed in the Appendix (Tables \ref{tab:fit_v} and \ref{tab:fit_vlsr}).

\subsubsection{The solar neighborhood}\label{res_sun}

To compare our results with previous works we focus now on the statistical properties of NSs in the so-called solar region, $7.5\,\leq\,R\,\leq\,9.5$ kpc, as in BR, BM and Z95.

From the fits we discussed in the previous subsection, we obtain the local surface density $\Sigma_{0}$ which varies from $\sim\,0.4\,-\,2\,\times\,10^{5}\,(N_{star}/10^{9})\,\rm{kpc}^{-2}$.
The volume density in the solar vicinity, $n_{0}$, also varies by a factor 5 between models, from $\sim\,1\,-\,5\,\times\,10^{-4}$ $(N_{star}/10^{9})\,\rm{pc}^{-3}$.

We can now infer the distance of the nearest NS simply by computing the minimum volume around the Sun which contains at least a NS, assuming constant density

\begin{equation}
	 1 = \frac{4\pi}{3}\,n_{0}\,d^{3}\, {\Rightarrow}\, d_{min} = \Big( \frac{3}{4\,\pi\,n_{0}} \Big)^{1/3};
\end{equation}

\noindent typical values of $d_{min}$ are around 10 pc.

\begin{table*}
	\centering
	\caption{Statistical properties of NSs at the solar circle.}		
	\begin{tabular}{c c c c c c c c c c c}
\hline
Model & 1A & 1B & 1C & 1D & 1E & 1A* & 1B* & 1C* & 1D* & 1E* \\
\hline
\\
$\Sigma_{0}$
& 0.44 & 1.02 & 0.90 & 1.21 & 1.93
& 0.63 & 1.31 & 1.17 & 1.55 & 2.33 \\
$[10^{5}\,(\frac{N_{star}}{10^{9}})\,\rm{pc}^{-2}]$ \\
\\
$n_{0}$
& 1.1 & 2.6 & 2.3 & 3.0 & 4.8
& 1.6 & 3.3 & 2.9 & 3.9 & 5.8 \\
$[10^{-4}\,(\frac{N_{star}}{10^{9}})\,\rm{pc}^{-3}]$ \\
\\
$d_{min}$
& 12.9 & 9.8 & 10.2 & 9.2 & 7.8
& 11.5 & 9.0 & 9.3 & 8.5 & 7.4 \\
$[\rm{pc}]$\\
\\
$<\!v\!>$
& 216 & 213 & 204 & 206 & 216
& 248 & 242 & 234 & 238 & 249 \\
$[\rm{km\,s}^{-1}]$ \\
\\
$<\!v^{LSR}\!>$
& 173 & 140 & 191 & 158 & 72
& 191 & 150 & 203 & 167 & 79 \\
$[\rm{km\,s}^{-1}]$ \\
\\
$f_{50}$
& 0.004 & 0.001 & 0.003 & 0.001 & $\ll\,0.001$
& 0.002 & $<\,0.001$ & 0.002 & $<\,0.001$ & $\ll\,0.001$\\
\\
$f_{50}^{LSR}$
& 0.025 & 0.043 & 0.035 & 0.084 & 0.497
& 0.014 & 0.032 & 0.034 & 0.078 & 0.454 \\
\\
\hline
	\end{tabular}
		\label{tab:sun}
\end{table*}

\subsection{Halo neutron stars}

\begin{figure}
	\centering
		\includegraphics[width=0.5\textwidth]{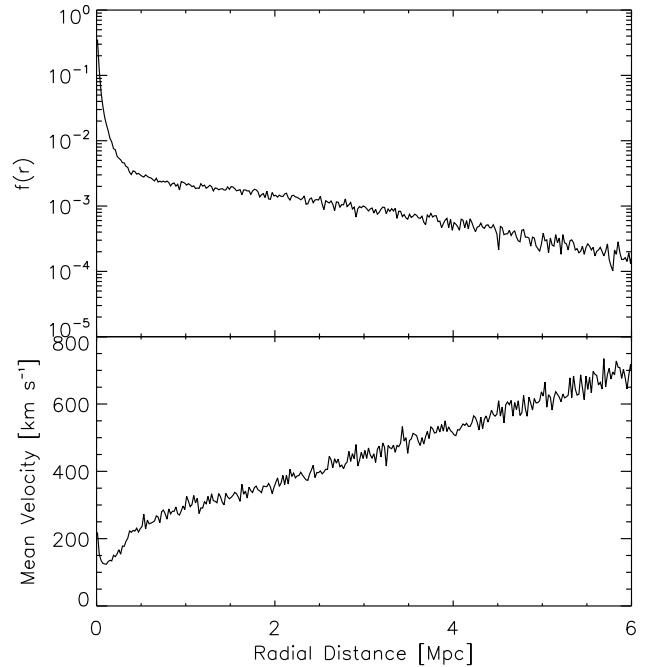}
	\caption{Model 1A - (upper panel) Radial distribution of NSs in the halo. (lower panel) Mean velocity.}
	\label{fig:halo}
\end{figure}

The distribution of NSs in the halo is shown in Fig. \ref{fig:halo}.
Bound NSs can be found as far as $\sim$ 1.5 Mpc, however the majority lies within the virial radius of the MW ($\sim\,270\,\rm{kpc}$).
The radial distribution clearly shows that unbound NSs start to be dominant at $\sim$ 500 kpc.
Accordingly the mean velocity, after an initial drop, starts to rise almost linearly from $\sim$ 500 kpc.
The gravitational effects of other galaxies (e.g. Andromeda and the Magellanic Clouds) have not been considered.

\subsection{Density maps}

\begin{figure}
	\centering
		\includegraphics[width=0.47\textwidth]{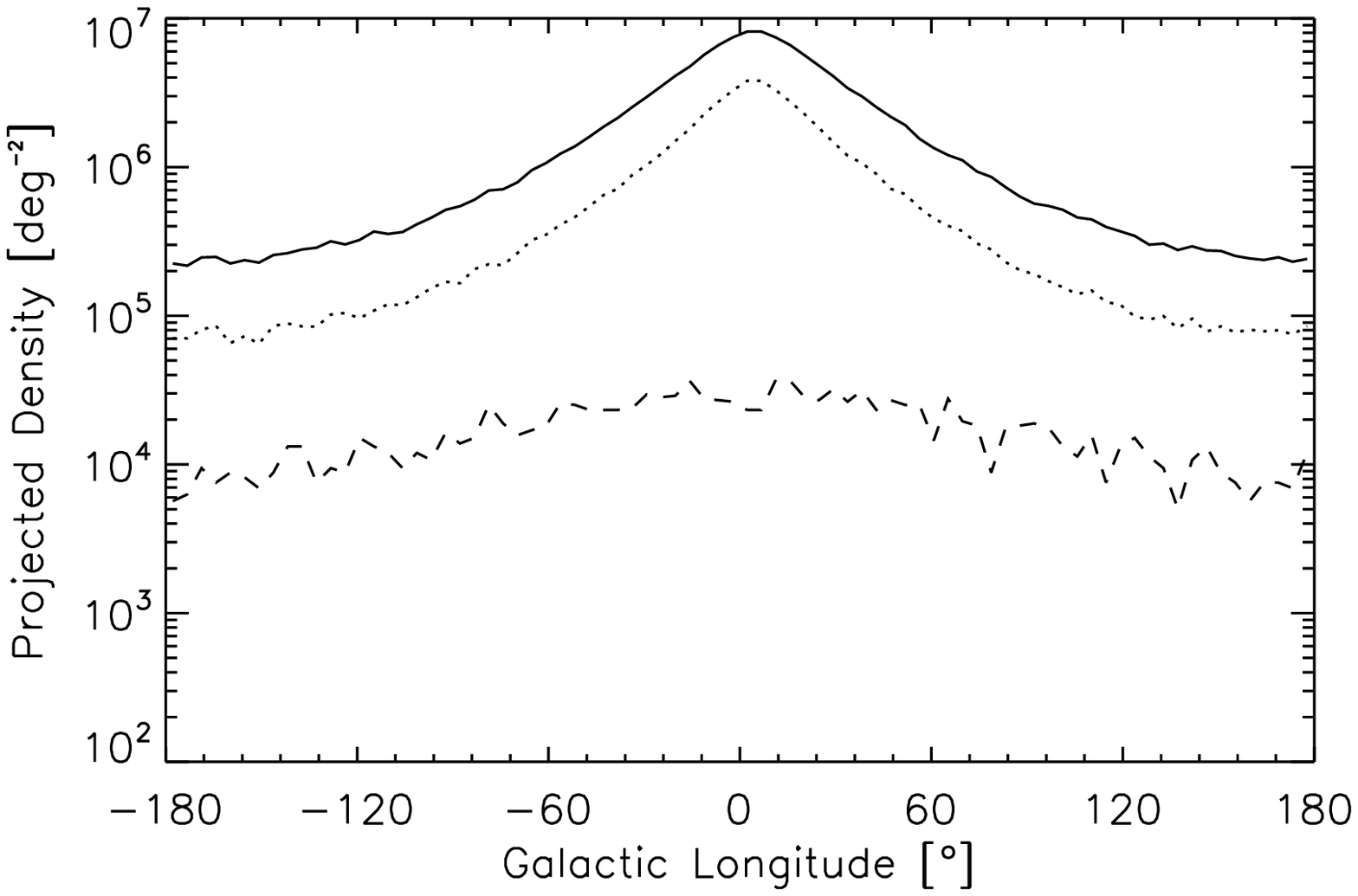}
		\includegraphics[width=0.47\textwidth]{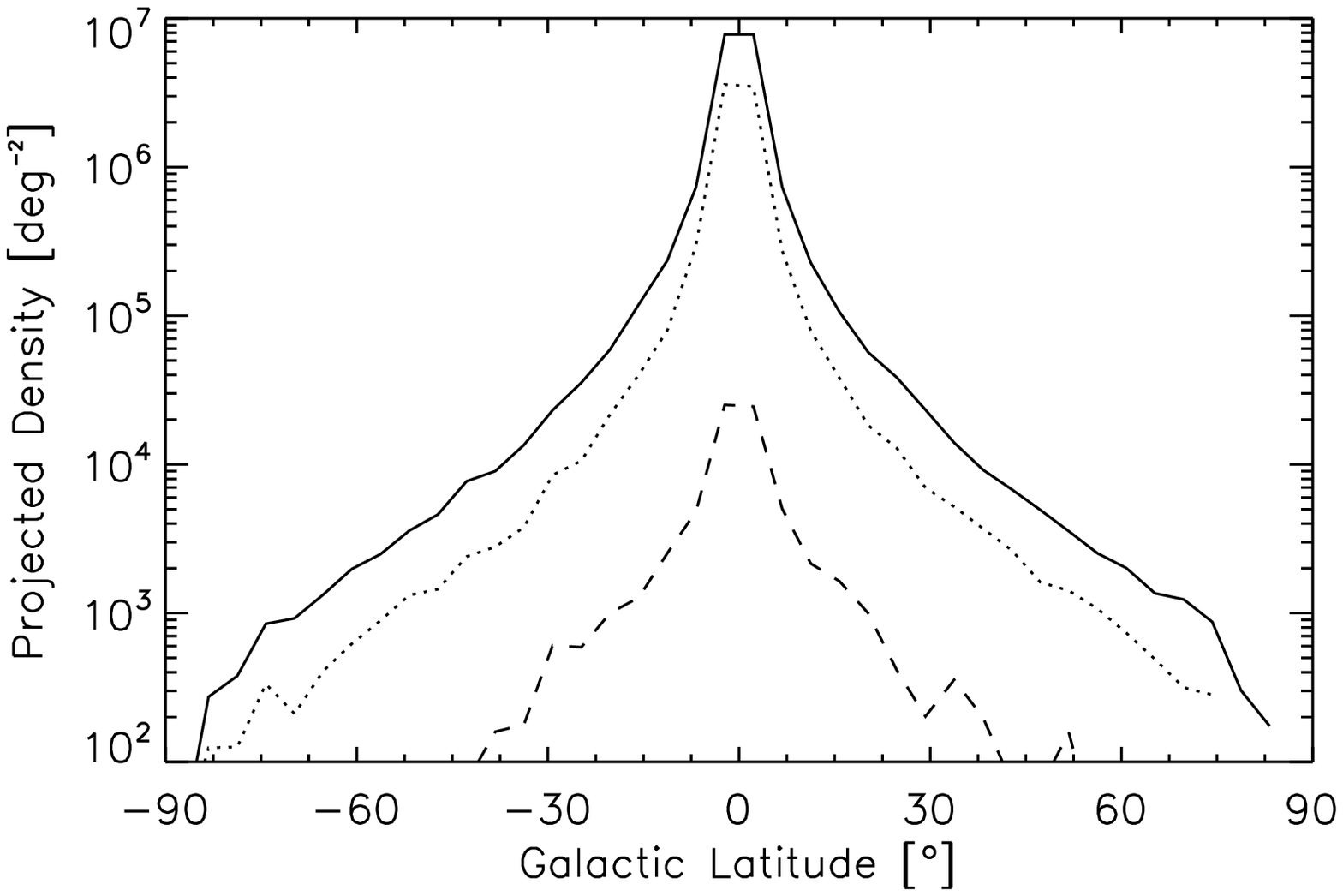}
	\caption{Density profile on the Galactic plane ($b=0$, upper panel) and along a meridian ($l=0$, lower panel) for $N_{star}=10^{9}$ - Model 1C. Cut-off distances are 30 (solid line), 10 (dotted) and 3 (dashed) kpc respectively.}
	\label{fig:ltrend}
\end{figure}

\begin{figure}
	\centering
		\includegraphics[width=0.47\textwidth]{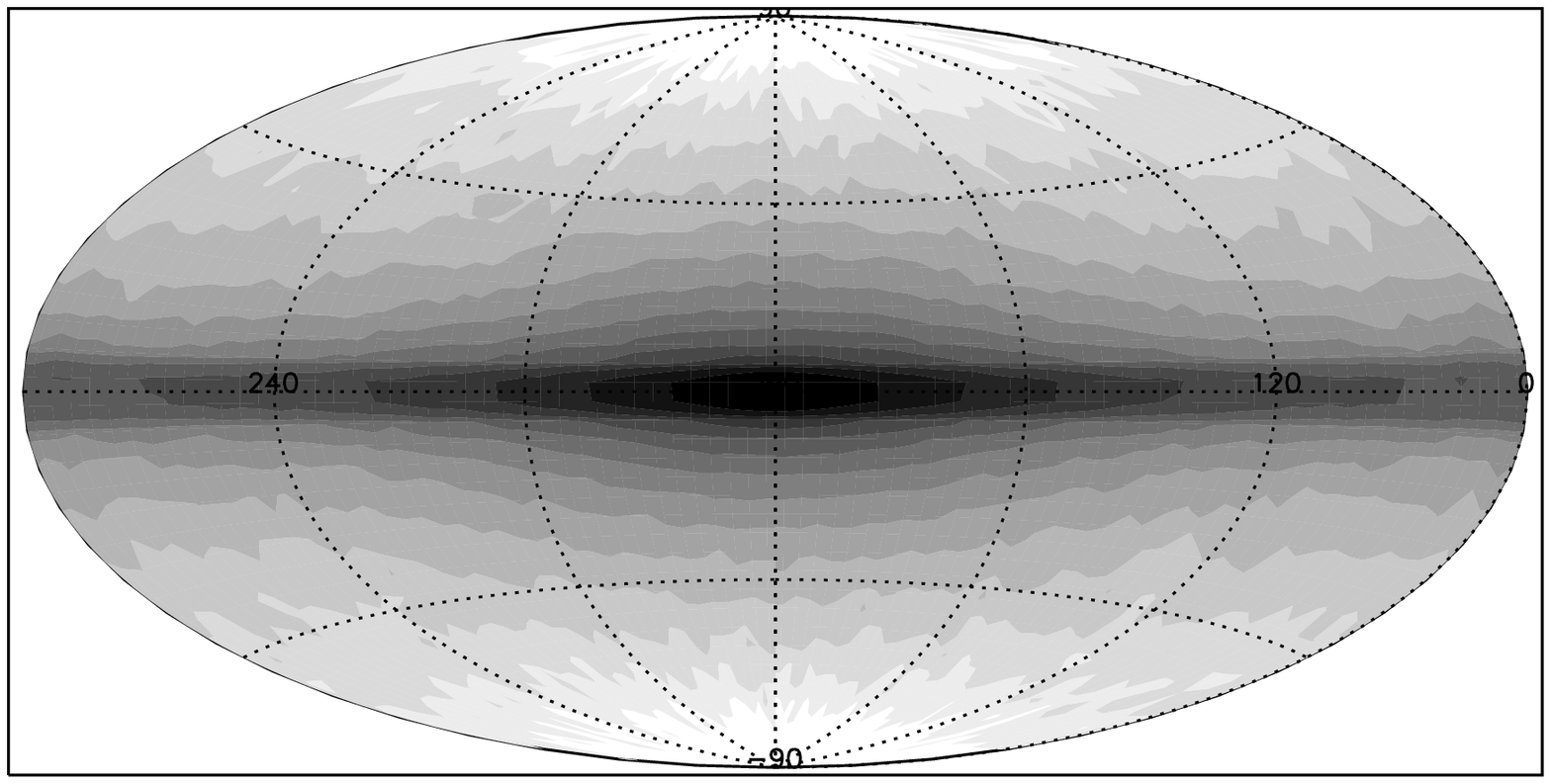}
		\includegraphics[width=0.47\textwidth]{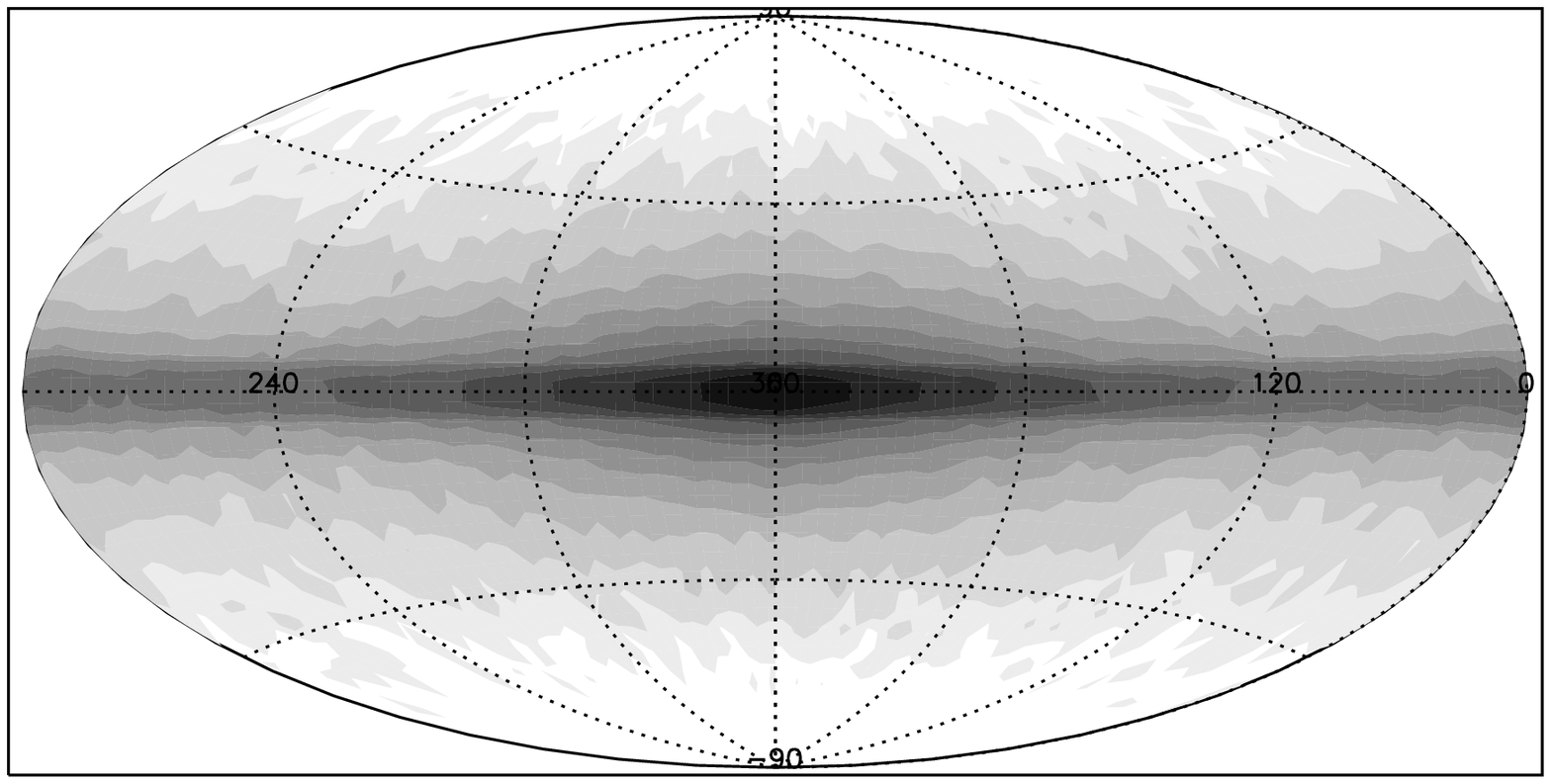}
		\includegraphics[width=0.47\textwidth]{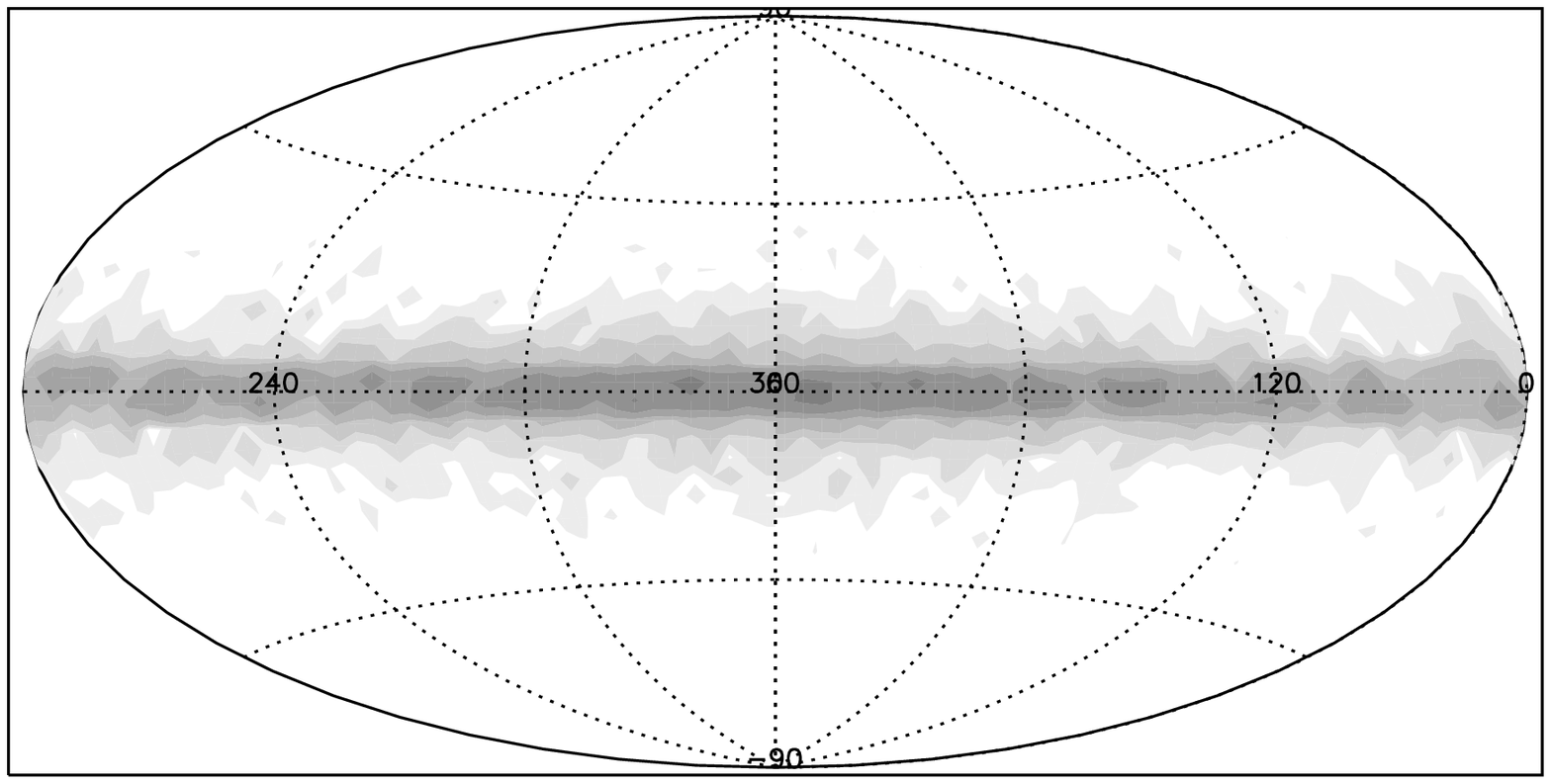}
	\caption{Sky maps of the projected density ($N_{star}=10^{9}$) - Model 1C. The cut-off distances are 30 kpc (upper panel), 10 kpc (central panel) and 3 kpc (lower panel) respectively.
	The density scale is normalized to the maximum density at 30 kpc.}
	\label{fig:skymaps}
\end{figure}

We compute the projected number density of NSs in heliocentric coordinates (\textit{l}, \textit{b}, \textit{d}) and give the relative sky maps for stars within 30, 10 and 3 kpc respectively.
Our sky maps (Fig. \ref{fig:skymaps}) clearly show that the most promising direction to look for NSs is towards the Galactic center, where the density is higher.
Moving away from the center, the density drops rapidly even along the Galactic plane (Fig.\ref{fig:ltrend}).

In Table \ref{tab:skydens} we list the inferred values of the projected density towards specific lines of sight (LOS).
Tha sampling distance varies according to the LOS: for example, towards the GC the sampling distance, $d_{max}$, is equal to $R_{0}$ while for the 3 other LOS lying on the Galactic plane $d_{max}$ is equal to the distance at which the LOS itself crosses the outer border of the stellar disk (20 kpc).

For large values of $d_{max}$ the projected density has non-negligible values even at high Galactic latitudes.
One intriguing consequence is that NSs in the halo may contribute to the observable rate of microlensing events, both of Galactic and extragalactic sources (Sartore et al., in preparation).
We thus calculate also the expected number density of NSs in the direction of the Magellanic Clouds, assuming 48 and 61 kpc for distance of the Large and Small Magellanic Clouds respectively (Table \ref{tab:skydens}).

\begin{table*}
	\centering
	\caption{Projected density of NSs towars different lines of sight.}
	\begin{tabular}{c c c c c c c c c c c}
\hline
\\
Line of sight & \multicolumn{10}{c}{Density $[(\frac{N_{star}}{10^{9}})\,\rm{deg}^{-2}]$} \\
\\
 & 1A & 1B & 1C & 1D & 1E & 1A* & 1B* & 1C* & 1D* & 1E* \\
\hline
\\
$l=0°,\,b=0°$
& 0.42 & 0.90 & 0.84 & 1.01 & 1.08
& 0.61 & 1.16 & 1.05 & 1.22 &  1.26 \\
$[\times\,10^{4}]$ \\
\\
$l=90°,\,b=0°$
& 0.80 & 1.64 & 1.22 & 1.51 & 2.40
& 1.18 & 2.05 & 1.53 & 1.85 & 2.76 \\
$[\times\,10^{3}]$ \\
\\
$l=180°,\,b=0°$
& 1.74 & 3.29 & 2.72 & 3.46 & 5.12
& 2.49 & 4.30 & 3.52 & 4.07 & 5.96 \\
$[\times\,10^{2}]$ \\
\\
$l=270°,\,b=0°$
& 0.82 & 1.62 & 1.22 & 1.53 & 2.37
& 1.16 & 2.04 & 1.51 & 1.89 & 2.75 \\
$[\times\,10^{3}]$ \\
\\
$b=+90°$
& 5.12 & 2.83 & 2.86 & 3.55 & 3.63
& 0.76 & 2.84 & 4.39 & 6.47 & 1.48 \\
$[\times\,10]$ \\
\\
$b=-90°$
& 2.13 & 4.25 & 4.26 & 5.73 & 2.18
& 4.55 & 4.97 & 1.46 & 2.16 & 2.96 \\
$[\times\,10]$ \\
\\
$l=280°,\,b=-33°$
& 3.25 & 3.35 & 3.47 & 2.87 & 2.46
& 3.79 & 3.39 & 4.32 & 3.31 & 2.66 \\
$[\times\,10^{2}]$ \\
\\
$l=303°,\,b=-44°$
& 4.44 & 4.25 & 3.76 & 2.78 & 2.78
& 5.20 & 3.85 & 4.07 & 3.02 & 2.79 \\
$[\times\,10^{2}]$ \\
\\
\hline
		\end{tabular}
	\label{tab:skydens}
\end{table*}

\subsection{Results with different potential}

Calculations made with the updated potential are labelled with an asterisk in the tables and give the following results.
The retention fraction $f_{bound}$ exhibits a significant increase, especially for models 1A* and 1C*, while for the remaining ones the enhancement is less conspicuous.
In all cases the fraction of NSs retained by the disk is only slightly increased (Table \ref{tab:disk}).

In all models with the updated potential, the scale height of the population is lower due to the increased masses of the disk and bulge, the decrease in $z_{1/2}$ being of the order of 10 - 20 percent (see the half density half thickness in Table \ref{tab:disk}).

The higher number of NSs retained by the disk implies also higher values of the surface and volume densities, together with the projected density for LOSs in the Galactic plane.
For the same reason, even relatively fast NSs can be found in the disk, thus increasing the mean velocities on NSs (see Tables \ref{tab:disk} and \ref{tab:sun}).

\section{DISCUSSION}\label{discussion}

Our calculations show that the distribution of birth velocities is the main factor driving the dynamics of NSs in the MW.
We obtain substantially different values of $f_{bound}$ among different birth velocity models, with the shape of the distribution (position of the peak, bimodality, etc.) also playing a role in determining the final fate of bound NSs.

The highest escape fraction, $\sim\,0.3$, are obtained with H05, F06E and F06L distributions.
This value is lower than those found by \cite{LL} (hereafter LL) and A02 and similar to that inferred by H05.
This is probably due to the fact that the velocity distributions proposed by LL and A02 were obtained adopting the distribution of free electrons of \cite{TC93}.
Both H05 and F06 adopted instead the revised model of \cite{CL02} for free electrons, which reduced estimates for distance, and thus velocity, of young pulsars.
We performed a run with the distribution of A02, obtaining $f_{bound}\sim\,0.54$, confirming their results on the escape fraction.

The adoption of the updated potential (higher mass of the Galaxy) implies higher escape velocities and hence only the fastest NSs, $\sim\,10 - 15$ percent, can definitely escape from the MW.

Albeit more than 70 percent of the NSs born in the MW are in bound orbits, the present-day number of NSs in the disk is only a small fraction of the total, $\lesssim 0.20$. The remaining ones are found in the halo where they spend most of their life.
This is a striking result but was not totally unexpected because, given their high spatial velocities, our synthetic NSs leave the disk in a short timescale, $\sim\,1\,-\,10$ Myr.
Another remarkable finding is that the ratio of young to old NSs in the disk is very low: for each neutron star detected as a young active source there should be still more than 100 old NSs hiding in the disk.

Our simulated NSs are born with significantly higher velocities with respect to what is found in other works.
In spite of this, our results for the half density half thickness show no significant differences with previous studies (except for the F06P distribution).
Also, the local spatial density of NSs falls between those found by BR, BM, Z95 and that of P90, i.e. approximately between 1 and $5\,\times\,10^{-4}\,(N_{star}/10^{9})\,\rm{pc}^{-3}$.
This means that the nearest neutron star lies within $\sim\,10$ pc from the Sun.

The mean velocity is higher by at least a factor $\sim\,2$ with respect to, for example, that found by BR, BM and Z95.
Low velocity NSs ($v\,\leq\,50\,\rm{km\,s}^{-1}$) in the disk are a tiny fraction, $f_{50}\,\sim\,0.001$.
This fraction grows by a factor $\sim 10$ in the LSR where $f_{50}^{LSR}$ is $\sim\,0.05$.
Again results obtained with the F06P distribution show a rather different behavior, with $f_{50}\,\ll\,0.01$ in the GRF while in the LRF roughly half of NSs in the disk are in the low velocity tail (Tables \ref{tab:disk} and \ref{tab:sun}).
However, the effective weight of the low velocity tail of the distribution of birth velocities is still matter for debate.

Most of NSs, both bound and unbound, run away from the Galactic plane in a short timescale and form a halo which extends well beyond the virial radius of the MW.
The phase-space distribution of halo NSs clearly shows a separation between bound and unbound NSs.
Unbound NSs become dominant at $r\sim\,500$ kpc.

The results presented in this paper will enable us to revisit a number of problems concerning isolated old NSs, like the accretion luminosity and its observability, the strategies for observing very close NSs, say within 100 pc and the optical depth of NSs in the perspective of using gravitational lensing to probe the population.

\begin{acknowledgements}
We thank the anonymous referee for several helpful comments which improved the previous versions of this paper.
We thank also M. C. Smith for helpful suggestions on the parameters of the Milky Way potential.
NS wishes to thank R. Salvaterra for useful comments on the manuscript and L. Paredi for technical support. The work of RT is partially supported by INAF/ASI through grant AAE-I/088/06/0.
\end{acknowledgements}


\appendix

\section{Coefficients of the fits.}\label{append_fits}

We give here the best fit parameters for the surface density of the disk (Table \ref{tab:fit_surf}), the height distribution (Table \ref{tab:fit_z}) and the cumulative velocity distributions in the disk, both in the GRF (Table \ref{tab:fit_v}) and in the LSR (Table \ref{tab:fit_vlsr}).

\begin{table*}
	\centering	
		\caption{Surface density of the disk.}
	\begin{tabular}{c c c c c c c c c c c c}
\hline
Model & 1A & 1B & 1C & 1D & 1E & 1A* & 1B* & 1C* & 1D* & 1E*\\
\hline
\\
$a_{0}$
& 6.09 & 6.48 & 6.51 & 6.60 & 6.54
& 6.24 & 6.57 & 6.57 & 6.66 & 6.61 \\
\\
$a_{1}$
& -2.54 & -2.58 & -3.08 & -2.79 & -1.74
& -2.47 & -2.41 & -2.84 & -2.68 & -1.83 \\
$[\times10^{-1}]$ \\
\\
$a_{2}$
& 1.40 & 1.57 & 2.37 & 1.83 & 0.32
& 1.05 & 1.20 & 1.96 & 1.79 & 0.58 \\
$[\times10^{-2}]$ \\
\\
$a_{3}$
& -5.69 & -8.06 & -12.70 & -9.20 & -2.28
& -2.99 & -5.25 & -10.22 & -10.07 & 0.13 \\
$[\times10^{-4}]$ \\
\\
$a_{4}$
& 0.90 & 1.67 & 2.59 & 1.82 & -0.34
& 0.22 & 1.00 & 2.11 & 2.25 & 0.34 \\
$[\times10^{-5}]$ \\
\\
Error
& 4 & 3 & 3 & 3 & 2
& 5 & 3 & 2 & 3 & 2 \\
$[\%]$ \\
\hline
		\end{tabular}
		\label{tab:fit_surf}
\end{table*}

\begin{table*}
	\centering	
	\caption{Distribution of heights.}
	\begin{tabular}{c c c c c c c c c c c}
\hline
Model & 1A & 1B & 1C & 1D & 1E & 1A* & 1B* & 1C* & 1D* & 1E* \\
\hline
\\
$b_{0}$
& 142.3 & 87.5 & 123.1 & 92.9 & 112.7
& 122.7 & 73.7 & 115.5 & 91.5 & 107.4 \\
\\
$b_{1}$
& 1.12 & 1.20 & 1.16 & 1.22 & 1.22
& 1.14 & 1.25 & 1.17 & 1.24 & 1.24 \\
\\
$b_{2}$
& -136.3 & -83.7 & -120.0 & -91.0 & -111.9
& -117.1 & -70.6 & -112.7 & -89.9 & -106.8 \\
\\
Error
& 20 & 41 & 50 & 63 & 48
& 22 & 46 & 63 & 65 & 56 \\
$[\%]$ \\
\hline
		\end{tabular}
		\label{tab:fit_z}
\end{table*}

\begin{table*}
	\centering	
	\caption{Cumulative velocity distribution in the disk.}
	\begin{tabular}{c c c c c c c c c c c}
\hline
Model & 1A & 1B & 1C & 1D & 1E & 1A* & 1B* & 1C* & 1D* & 1E* \\
\hline
\\
$v_{0}$
& 214.1 & 207.4 & 201.1 & 200.1 & 207.2
& 244.5 & 241.2 & 232.5 & 232.5 & 238.6 \\
$[\rm{km\,s}^{-1}]$ \\
\\
\textit{n}
& 3.98 & 4.56 & 4.05 & 4.63 & 8.09
& 4.03 & 4.84 & 4.16 & 4.85 & 8.21 \\
\\
\textit{m}
& 3.98 & 4.56 & 4.05 & 4.63 & 8.09
& 4.03 & 4.84 & 4.16 & 4.85 & 8.21 \\
\\
Error
& 72 & 91 & 91 & 95 & 100
& 63 & 87 & 93 & 97 & 100 \\
$[\%]$ \\
\hline
		\end{tabular}
		\label{tab:fit_v}
\end{table*}

\begin{table*}
	\centering
	\caption{Cumulative velocity distribution in the disk (LSR).}
	\begin{tabular}{c c c c c c c c c c c}
\hline
Model & 1A & 1B & 1C & 1D & 1E & 1A* & 1B* & 1C* & 1D* & 1E* \\
\hline
\\
$v'_{0}$
& 162.4 & 130.5 & 169.4 & 133.5 & 57.1
& 178.7 & 139.7 & 182.3 & 141.9 & 64.4 \\
$[\rm{km\,s}^{-1}]$ \\
\\
$n'$
& 3.35 & 3.33 & 2.77 & 2.57 & 1.92
& 3.34 & 3.35 & 2.70 & 2.50 & 1.94 \\
\\
$m'$
& 2.35 & 3.33 & 2.76 & 2.57 & 1.92
& 3.34 & 3.35 & 2.70 & 2.50 & 1.94 \\
\\
Error
& 73 & 70 & 44 & 18 & 26
& 67 & 62 & 43 & 12 & 26 \\
$[\%]$ \\
\hline
		\end{tabular}
		\label{tab:fit_vlsr}
\end{table*}

\end{document}